# Super-resolution enhancement by quantum image scanning microscopy

Ron Tenne[1*], Uri Rossman[1*], Batel Rephael[1*], Yonatan Israel[1,2], Alexander Krupinski-Ptaszek[3], Radek Lapkiewicz[3], Yaron Silberberg[1], Dan Oron[1]

[1] - Department of Physics of Complex Systems, Weizmann Institute of Science, Rehovot 76100, Israel

[2] - Department of Physics, Stanford University, Stanford, CA 94305, USA

[3] - Institute of Experimental Physics, Faculty of Physics, University of Warsaw, Pasteura 5, 02-093 Warsaw, Poland

[*] - These authors contributed equally to this work

Correspondence and requests for materials should be addressed to D.O. (email: dan.oron@weizmann.ac.il)

## Abstract

The principles of quantum optics have yielded a plethora of ideas to surpass the classical limitations of sensitivity and resolution in optical microscopy. While some ideas have been applied in proof-of-principle experiments, imaging a biological sample has remained challenging mainly due to the inherently weak signal measured and the fragility of quantum states of light. In principle, however, these quantum protocols can add new information without sacrificing the classical information and can therefore enhance the capabilities of existing super-resolution techniques. Image scanning microscopy (ISM), a recent addition to the family of super-resolution methods, generates a robust resolution enhancement without sacrificing the signal level. Here we introduce quantum image scanning microscopy (Q-ISM): combining ISM with the measurement of quantum photon correlation allows increasing the resolution of ISM up to two-fold, four times beyond the diffraction limit. We introduce the Q-ISM principle and obtain super-resolved optical images of a biological sample stained with fluorescent quantum dots using photon antibunching, a quantum effect, as a resolution enhancing contrast mechanism.

## Main Text

The diffraction limit, as formulated by Abbe, sets the attainable resolution in far-field optical microscopy to about half of the visible wavelength[1], hindering its applicability in life science studies at very small scales. Over the past two decades, several super-resolution methods have successfully overcome the diffraction limit, including emission depletion microscopy, localization microscopy and structured illumination microscopy[2–6]. The continuous and rapid improvement in detector technology has enabled two more recent developments in the field of super-resolution microscopy, which are the center of this work: quantum super-resolution microscopy and image scanning microscopy (ISM). As for the first, a surge of interest in super-resolution imaging based on quantum optics concepts[7–13], inspired and facilitated by the progress in high temporal resolution imagers, resulted in a few successful proof-of-principle demonstrations[7,8,14]. The second, ISM, relies on a small array of fast detector and offers a two-fold enhancement of resolution[15,16]. Since ISM is compatible with a standard confocal microscope architecture it has already been integrated into commercial products.

While all super-resolution modalities violate at least one of the basic assumptions of the Abbe theory, many rely on breaking more than one. For instance, stimulated emission depletion (STED) and saturated structured illumination microscopy (SSIM) breach both the assumption of a linear response of a fluorophore to the excitation light and that of a uniform illumination field[17,18]. In contrast, the few demonstrations of quantum super-resolution microscopy[7,8,14] relied solely on violating the implicit assumption, underlying Abbe's derivation, that light behaves as waves rather than particles. ISM, as well, depends on violating a single assumption, a uniform illumination of the sample.

An alternative approach to take advantage of quantum properties of light to break Abbe's limit is to enhance an already established super-resolution method with the extra information held in a photon correlation measurement[12]. ISM is a natural candidate for this purpose since it does not compromise the collected signal level. In ISM each pixel in a detector array acts as a small pinhole in a laser-scanning confocal microscope (LSCM) recording an image with a resolution twice as fine as the diffraction limit[16,19]. Thus, one can construct a super-resolved image without reducing the collected signal level[20,21]. Some variants offer the same resolution improvement within an all-optical setup reducing the need in computational power and fast imagers[22–24]. Although modest, this resolution improvement is robust to changes in the sample and label type and can be integrated with additional microscopy modalities[25,26].

We present here a new super-resolution scheme, quantum image scanning microscopy (Q-ISM), utilizing the measurement of quantum correlations in an ISM architecture. During a confocal scan, each pair of detectors in a detector array generates a sharp image using photon correlations. These images are merged together to surpass the resolution of both ISM and photon correlation measurements separately. By violating both the classical light and uniform illumination assumptions, we were able to obtain super-resolved images of a biological sample based on a quantum optical effect, namely photon antibunching.

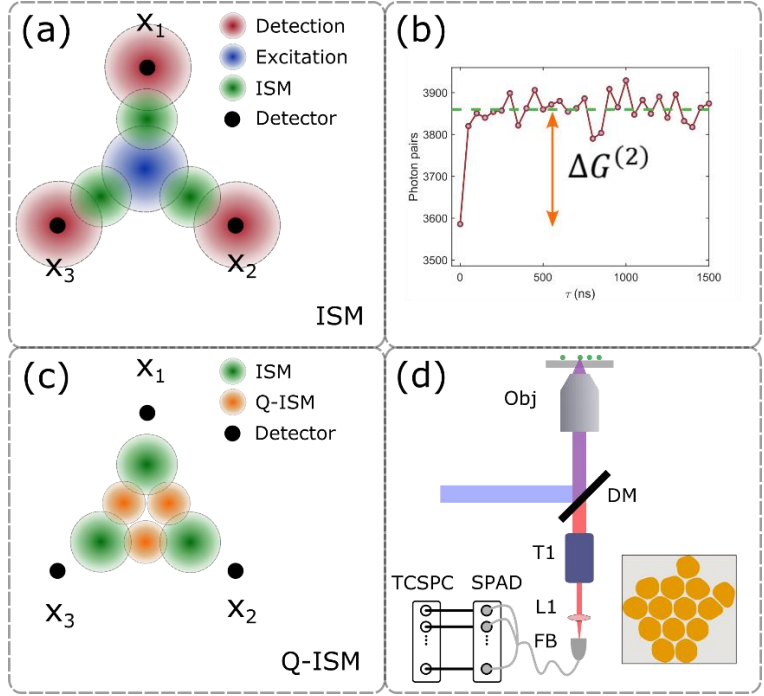

**Fig. 1| Q-ISM principle of operation. (a)** Schematic of the ISM method. The ISM PSF for each detector (green circles) is a product of the excitation laser beam profile (blue circle) and the detection probability distribution (red circles) centered around the detector position (black circles). **(b)** Second order correlation function, $G^{(2)}(\tau)$, taken from a single position within a scan and binned according to the laser pulse period ($50\ ns$). The dashed green line shows the average number of photon pairs measured at long time delays. The orange arrow indicates the number of missing photon pair events, the contrast of Q-ISM **(c)** Schematic of Q-ISM. The effective PSF for each detector pair (orange circles) is a product of the two ISM PSFs of the two detectors (green circles). **(d)** Optical setup: Obj – objective lens, DM – dichroic mirror, T1 – variable telescope, L1 – lens, FB – fiber bundle, SPAD – single-photon avalanche detectors, TCSPC – time-correlated single photon counting module. The inset shows a segmented image of the fiber-bundle's entrance facet highlighting the 14 fibers used in this work.

### Resolution improvement in Q-ISM

To explore the resolution gain in Q-ISM let us consider the case of $M$ identical emitters positioned at $x_i$ ($i = 1..M$) and imaged by a unity magnification imaging system whose incoherent point spread function (PSF) is $h(x)$. Applying a uniform illumination field yields an intensity image

$$G^{(1)}(x) \propto \sum_{i=1}^{M} h(x - x_i). \quad (1)$$

In standard ISM, presented schematically in figure **1**a, the sample is illuminated with a laser beam focused to a (diffraction limited) spot centered around $x = 0$ (blue circle)[21]. As the sample position is scanned, the photoluminescence signal is recorded by an array of point-like detectors

(full black circles), whose image plane positions are denoted by $\bar{x}_\alpha$ ($\alpha = 1, .., K$). Neglecting the fluorescence Stokes shift, a single detector $\alpha$ obtains an intensity image

$$G_\alpha^{(1)}(x) \propto \sum_{i=1}^{M} h(x_i - x) \cdot h(\bar{x}_\alpha - (x_i - x)), \tag{2}$$

where $x$ is the scan coordinate. The product of the two terms in equation **Error! Reference source not found.** is depicted in figure **1a**; the multiplication of the excitation probability distribution (blue circle) and the detection probability distribution of each detector (red circles) results in a narrower effective PSF (green circles) centered around $\frac{1}{2}\bar{x}_\alpha$. The images from the $K$ different detectors can be shifted and summed into a single image $G_{ISM}^{(1)}(x) = \sum_{\alpha=1}^{K} G_\alpha^{(1)}(x - \frac{1}{2}\bar{x}_\alpha)$.

Although $G_{ISM}^{(1)}(x)$ contains spatial frequencies two times larger than the maximal spatial frequency contained in a widefield image, its resolution is enhanced only by a factor of $\sqrt{2}$. To extend the image resolution to its Fourier limit ones need to perform Fourier reweighting (FR)[16]. By digitally amplifying the high spatial frequency content, one obtains an image with an up to two-fold resolution enhancement and a signal-to-noise ratio (SNR) comparable to that of $G^{(1)}(x)$ (Supplementary Section 2).

Light emitted from fluorophores, such as dye molecules, quantum dots (QDs) and solid-state defects, exhibits photon antibunching[27–31], a well-known effect in quantum optics. Each such fluorophore emits at most a single photon for every excitation pulse that is much shorter than the fluorescence lifetime. Therefore, as presented in figure **1b**, the number of simultaneously detected photon pairs ($\tau = 0$) is lower than the number of photon pairs at a delay of one or more pulses ($\tau \neq 0$). The spatial distribution of missing events of simultaneously detected photon pairs measured by a pair of detectors positioned at $\bar{x}_\alpha$ and $\bar{x}_\beta$ yields an image of the photon correlation contrast following (see Supplementary Section 1)

$$\Delta G_{\alpha\beta}^{(2)}(x) \propto \sum_{i=1}^{M} [h(x_i - x)]^2 \cdot h(\bar{x}_\alpha - (x_i - x)) \cdot h(\bar{x}_\beta - (x_i - x)). \tag{3}$$

The effective two photon absorption and two photon detection contrast results in a PSF, schematically illustrated in figure **1c** (orange circles), which is a product of two effective ISM PSFs centered around $\frac{1}{2}\bar{x}_\alpha$ and $\frac{1}{2}\bar{x}_\beta$ (green circles). The photon correlation image is a sum of effective PSFs that are approximately centered at $x_i + \frac{1}{4}(\bar{x}_\alpha + \bar{x}_\beta)$ and are narrower by a factor of 2 compared to the original PSF. Similar to ISM, the $\frac{K(K-1)}{2}$ images of photon correlations can be shifted and summed to get the merged Q-ISM image,

$$G_{Q-ISM}^{(2)}(x) = \sum_{\alpha=1}^{K} \sum_{\beta=\alpha+1}^{K} \Delta G_{\alpha\beta}^{(2)}\left(x - \frac{1}{4}(\bar{x}_\alpha + \bar{x}_\beta)\right). \tag{4}$$

As in the case of ISM, the resolution enhancement of $G^{(2)}_{Q-ISM}(x)$ (approximately two) can be extended further by performing FR. FR can increase the resolution of a high SNR Q-ISM image up to four times beyond the diffraction limit (Supplementary Section 2).

Demonstration of resolution enhancement in Q-ISM

To demonstrate the concept of Q-ISM we initially image a sample of QDs randomly dispersed on a substrate. The microscope setup, schematically shown in figure **1**d, consists of a standard confocal microscope whose pinhole is replaced with a honey-comb structured fiber bundle (see inset of figure **1**d). The fiber bundle fans out to separate fibers, 14 of which are individually coupled to a single-photon avalanche detector (SPAD) (SPCM-AQ4C, Perkin-Elmer) feeding a time-correlated single-photon counting (TCSPC) card (DPC230, Becker & Hickl). A pulsed laser beam (EPL-475, Edinburgh Instruments, $473\ nm$, $80\ ps$ pulse width, $20\ MHz$ repetition rate, $3\ \mu W$ power), focused through a high numerical aperture (NA) objective (Plan Apo Vc 100x, Nikon, NA 1.4), illuminates the sample as it is scanned in the lateral dimensions with a piezo stage (P-542.2SL XY, Physik Instrumente). The detection timing data collected during the scan is parsed and analyzed to generate the ISM and Q-ISM images (see Methods and Supplementary Section 3 for further details).

Figure **2** compares the CLSM, ISM, Q-ISM and Fourier reweighted Q-ISM analyses from a scan of a few QDs. The standard CLSM image (Fig. **2**a) was obtained by summing the intensities of all detectors, effectively treating the fiber bundle as a semi-open pinhole in a confocal microscope with an effective diameter of 0.6 Airy units (AU). Here we observe a single unresolved elongated patch containing fluctuations, likely to result from quantum dot blinking. By shifting the image obtained from each of the detectors, a process termed pixel reassignment[19] (Supplementary Section 4), and summing those together, we generate an ISM image (Fig. **2**b). In addition to the enhanced resolution, temporal fluctuations in fluorescence during the sample scan (see Fig. **2**a) are smoothed, becoming unnoticeable in ISM (see Fig. **2**b), due to the delay accrued between the shifted detector images.

To generate a Q-ISM image, we analyzed the same data set to calculate the second order photon correlation (see Fig. **1**b). These correlations are analyzed for each position along the scan using the photon arrival times in every pair of detectors (Supplementary Section 3). As described above, taking the difference between the coincident and delayed photon pair detections yields an antibunching contrast image for each detector pair, $\Delta G^{(2)}_{\alpha\beta}(x)$, which can be merged according to equation **Error! Reference source not found.** into a Q-ISM image, $G^{(2)}_{Q-ISM}(x)$, shown in figure **2**c (Supplementary Sections 3 and 4). Finally, FR is performed on the image (Fig. **2**d) to flatten the spatial frequency response of the imaging technique and bring the resolution closer to the theoretical, 4-fold, enhancement limit (Supplementary Section 2). To emphasize the resolution improvement, figure **2**e presents interpolated cross-sections from the four images in figure **2**a-d. A quantitative assessment of the resolution improvement (Supplementary Section 5) shows that the resolution, according to the Rayleigh criterion, improves from $460\ nm$ in the standard widefield microscope to $350\ nm$ in ISM, $260\ nm$ in Q-ISM and finally $204\ nm$ in the FR Q-ISM image. The resolution enhancement factor compared to the widefield image is 1.3, 1.8 and 2.3 for the ISM, Q-ISM and FR Q-ISM images respectively. The fact that both the ISM and Q-ISM images do not fulfil the theoretical enhancement limit of $\sqrt{2}$ and 2 respectively is probably due to the non-negligible detector size in our system (a de-magnified diameter of $\sim 0.25\lambda$). In FR Q-ISM the resolution enhancement is further limited by the finite SNR of the Q-ISM images.

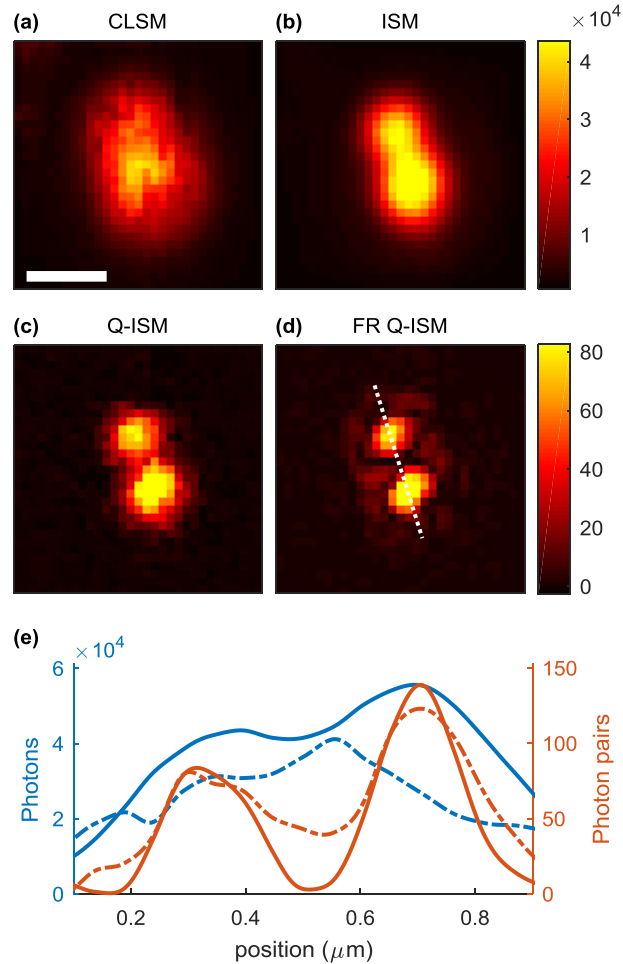

**Fig. 2| Resolving emitters with Q-ISM.** A $1.5\ \mu m$ by $1.5\ \mu m$ confocal scan ($50\ nm$ steps, $200\ ms$ pixel dwell time) of a sample of a cluster of CdSe/CdS/ZnS quantum dots. **(a)** The summed intensity over the whole detector array per scan position. **(b)** ISM image, $G^{(1)}_{ISM}(x)$: The scan image from each detector is shifted prior to summation. The color bar in (a) and (b) shows the number of detected photon counts. **(c)** Q-ISM image, $G^{(2)}_{Q-ISM}(x)$: the antibunching signal image of each detector pair is shifted and summed (see equation **Error! Reference source not found.**, and Supplementary Sections 1 and 3). **(d)** A Fourier reweighted (FR) image obtained from (c) (see Supplementary Section 2). The color bar in (c) and (d) shows the number of detected missing photon pairs. For clarity of

presentation, the maximal value of the FR image (d) was scaled to that of the Q-ISM image (c). **(e)** Interpolated cross-sections for the dotted white line in (d) for the four different analyses: CLSM (dashed blue), ISM (solid blue), Q-ISM (dashed red) and FR Q-ISM (solid red). Scale bar: $0.5 \mu m$.

## Quantum microscopy of a fixed cell sample

We further demonstrate the Q-ISM method by imaging a biological sample of micro-tubules, labeled with QDs, in fixed 3T3 cells (see Methods). In figure **3**, a super-resolved FR Q-ISM image of a $3 \mu m$ by $3 \mu m$ area is juxtaposed with the CLSM, ISM and Q-ISM images analyzed from the same data set. Although the SNR is lower in the Q-ISM and FR Q-ISM images (Fig. **3**c and **3**d), all the visible features in the CLSM (Fig. **3**a) and ISM (Fig.**3**b) images are present with a finer resolution. This observation demonstrates the valuable information contained in photon correlation data even though it is based on rare events. Supplementary Section 6 provides a more elaborate discussion regarding SNR in Q-ISM.

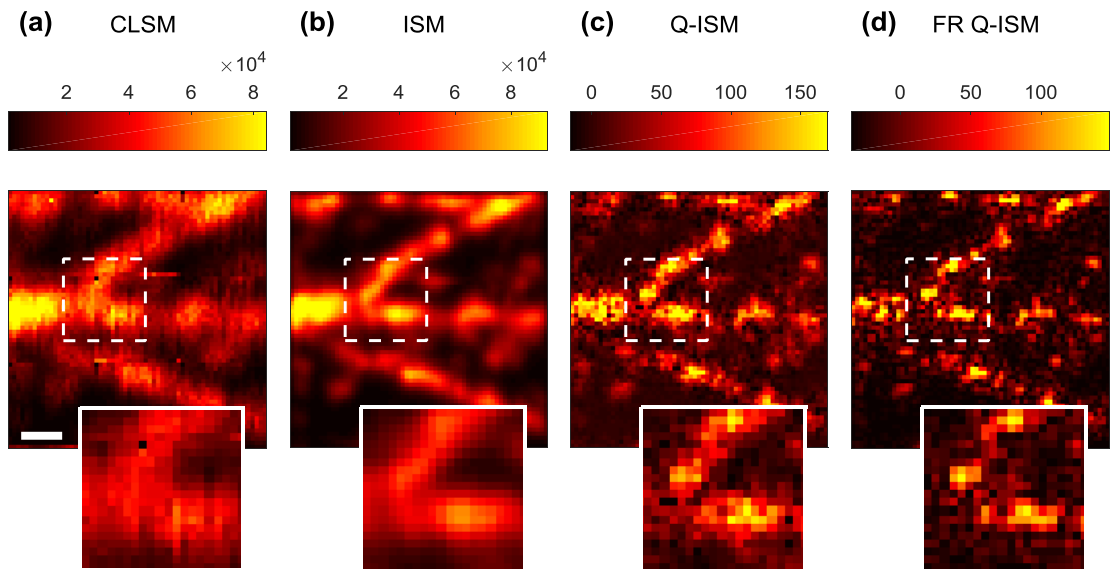

**Fig. 3 | Q-ISM of micro-tubule labeled cell samples.** Images analyzed from a confocal scan (50 $nm$ steps, 100 $ms$ pixel dwell time) of a 3 $\mu m$ by 3 $\mu m$ section of micro-tubules in a fixed 3T3 cell labeled with fluorescent quantum dots (QDot 625, Thermo Fisher). **(a)** CLSM image **(b)** ISM. **(c)** Q-ISM image. **(d)** A Fourier reweighted (FR) Q-ISM image. For clarity of presentation, the maximum of the FR image shown in (d) was scaled to the maximum of the Q-ISM image (c). A zoomed-in version of a section of the images, framed by the white dashed line, is shown below each image. Scale bar: 0.5 $\mu m$.

## Z-sectioning in Q-ISM

Q-ISM offers an improved resolution compared to ISM not only in the transverse directions but also in the axial direction. To explore the axial resolution enhancement, figure **4** presents images of a thin area from the same sample shown in figure **3**, for different imaging planes, shifted by 0.4 microns relative to each other. The objective defocusing results in widening of both the excitation and detection PSFs and therefore a decrease in the integrated signal level accompanies a blurring of the image. Figure **4**a-d show the ISM images obtained for objective positions $-0.8\ \mu m$, $-0.4\ \mu m$, 0 and $0.4\ \mu m$ respectively (0 designates the position of the collection optimal focus). As expected from a confocal microscope the fine image features are blurred and their total intensities decreases as we move from the focus position[19,20]. In comparison, the contrast of Q-ISM images, presented in figure **4**e-h, has a stronger dependence on defocusing; rejecting more of the out of focus contributions.

Heuristically, the sharper decrease in Q-ISM image intensity occurs since the contrast results from missing events in which two excitation photons are absorbed and two emitted photons are detected in a small detector array. The probability of those independent events occurring simultaneously decreases substantially with defocusing since both the excitation power density and the probability to detect over the small array decrease (Supplementary Section 7). This effect is highlighted in figure **4**i displaying the intensity sum for ISM (blue) and Q-ISM (red) over a thin section in the sample, *versus* the defocusing position. Although the improved axial resolution is clearly visible in figure **4**, the non-negligible sample thickness prevents a quantitative estimation of the resolution enhancement. To overcome this issue we analyze images of a thin sample of QDs spin coated on a cover slip obtained in similar experimental conditions (Supplementary Section 7). The full width half maxima (FWHM) of the integrated intensity in the case of Q-ISM ($0.8\ \mu m$) is narrower by a factor of ~2.4 than the FWHM for ISM ($1.8\ \mu m$).

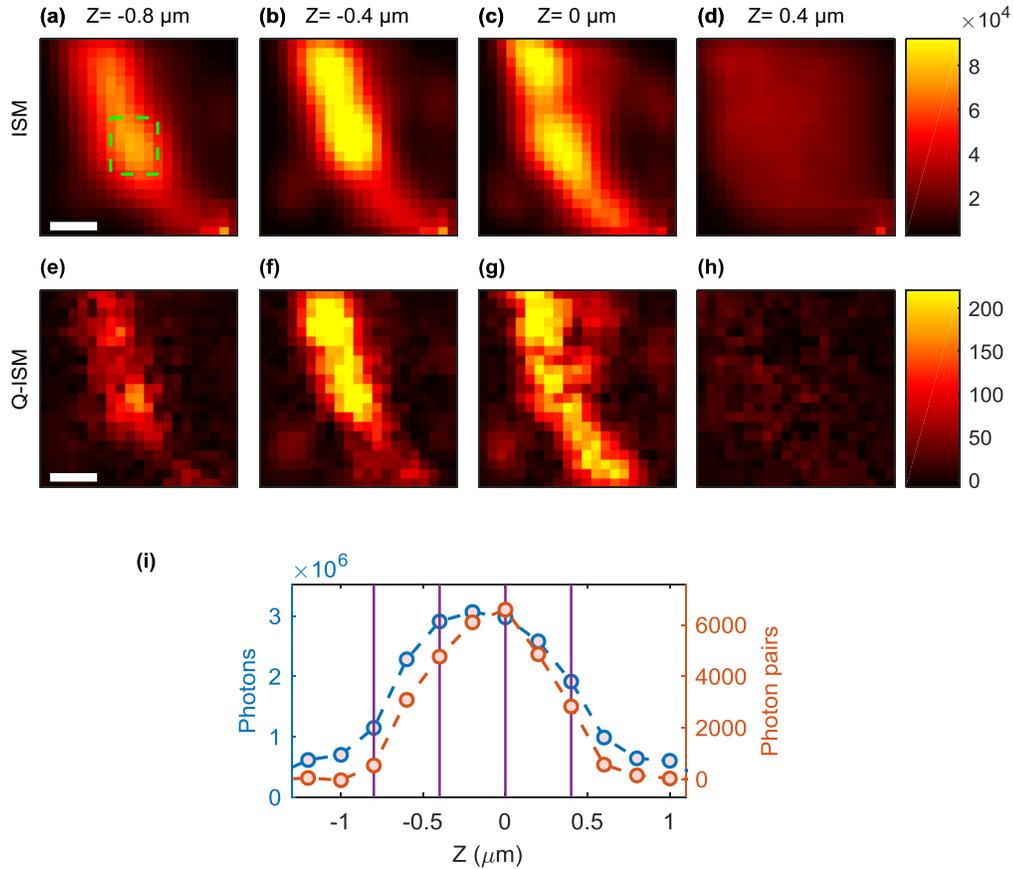

**Fig. 4| Resolving power of Q-ISM in the axial dimension. (a)-(d)** ISM images of a $1\mu m \times 1\mu m$ ($50 nm$ pixel size) axially thin region in the 3T3 cells sample at objective defocus position of -0.8 $\mu m$, -0.4 $\mu m$, 0 and 0.4 $\mu m$ respectively. 0 is the sample focus plain. **(e)-(h)** Q-ISM images analyzed from the same measurement as (a)-(d) respectively. **(i)** An integrated signal of the ISM and Q-ISM images over the area enclosed by the green dashed rectangle in (a) are shown in blue and red circles respectively. Violet vertical lines show the defocus positions of the images shown in panels (a)-(h). Scale bar for panels (a)-(h): 0.25 $\mu m$.

## Discussion

Q-ISM is compatible with standard confocal microscopy with a few additional requirements from the experimental setup and the fluorescent markers. First, the setup requires a small array of detectors with nanosecond scale temporal resolution such as a small monolithic SPAD array or a fiber bundle detector[32,33]. The latter setup avoids inter-detector cross-talk, a crucial advantage for accurate measurement of antibunching. In order to speed-up the acquisition process, conceptually similar ideas[11] may be implemented in a widefield configuration by employing a large area SPAD array[34]. Alternatively, using a large SPAD array, one can perform a multifocal adaptation[35] of the Q-ISM concept. As for the excitation source, either a pulsed or a continuous wave (CW) laser can be used without affecting the method's resolution or SNR. Here, we chose to work with a pulsed excitation source since it is favorable in avoiding photo bleaching under saturating illumination[36].

Relying on the photon correlation contrast poses some additional constraint on sample labeling. First, an isolated fluorescent probe should present some degree of antibunching. Notably, however, this is not a particularly limiting condition since commercial dye molecules, QDs, solid-state defects and some fluorescent proteins fulfill this requirement[31,37,38]. Since the photon correlation SNR increases monotonically with fluorescence quantum yield (QY) and the excitation repetition rate, labels should exhibit a high QY under close-to-saturation conditions. Photo-stability for a duration in the scale of a second, necessary for high SNR and to avoid scanning artifacts, has been demonstrated for colloidal QDs, solid-state defects and a variety of dye molecules routinely used in bio-imaging[31,36,38]. Finally, in order to achieve high photon rates under saturation conditions, the emitter's excited state lifetime should preferably be below a microsecond.

Unlike some of the previous implementations of imaging assisted by a photon correlation measurement[12], Q-ISM does not rely on spatial sparsity of emitters. Since each emitter adds to the occurrence of missing photon pairs, dense scenes, such as those presented in figure **3**, can be imaged. However, increasing the labeling density of the sample does not improve the SNR because both the number of missing photon pairs and its standard error grow linearly with the emitter density (Supplementary Section 6). In fact, given a magnification and pixel size, the saturation of the single-photon detectors sets an upper limit on the sample density and crossing that limit can degrade the image. In principle, the Q-ISM method could be extended to higher correlation orders to achieve an even higher resolution, but the SNR of correlation orders higher than two will decrease with labeling density (Supplementary Section 6). Therefore applying those as the contrast of ISM may prove challenging.

A few variants of ISM have already extended the lateral resolution or the z-sectioning capabilities of ISM by taking advantage of the sample's non-linear response. Gregor *et al.* applied a 2-photon excitation scheme to induce z-sectioning and gain a high penetration depth in an all-optical variant of ISM[26] whereas Laporte *et al.* harnessed the emitters' saturation to achieve higher lateral resolution[25]. Note that in both of these examples the excitation is non-linear while the detection process remains linear. In contrast, Q-ISM relies on missing photon pair events; events of absorption and detection of two photons. Therefore, the signal measured in Q-ISM depends not only on the excitation PSF squared but also on the second power of the detection PSF (see equation **Error! Reference source not found.**). The additional non-linearity in detection translates into higher available spatial frequencies and therefore a possibility for higher resolution, up to a 4-fold improvement over the diffraction limit (Supplementary Section 2). In that sense, Q-ISM realizes a theoretical concept, introduced by Hell *et al.* over two decades ago, of detecting only photon pairs with both photons emanating from the same position in a confocal setup[39,40].

Conclusions

In summary, Q-ISM combines the super-resolving capabilities of both quantum correlation measurements and ISM to achieve an up to four-fold enhanced resolution, significantly more than both methods separately. This combination enabled imaging of a biological sample, fixed cells labelled with single photon emitters, relying only on the antibunching contrast. Since Q-ISM provides information complementary to the conventional ISM data measured simultaneously, we believe that it may offer a substantial and practical improvement to ISM in terms of resolution and background. An algorithmic combination of the high resolution contained in a Q-ISM image with the high SNR provided by a standard ISM image may potentially lead to an image significantly surpassing each technique separately.

Methods

**Sample preparation.** CdSe/CdS/ZnS QDs were prepared in a colloidal synthesis. Details regarding the synthesis can be found in the supplementary information of reference [11]. The fluorescence was centered around $617\ nm$ and presented a $0.62$ quantum yield for a $530\ nm$ excitation wavelength. The QDs were diluted in a 3%wt solution of poly methyl methacrylate (PMMA) in Toluene and spin coated onto a microscope cover slip.

Fixed cell samples were prepared as follows. NIH 3T3 cells were grown to 80% confluency and fixed by 15 min incubation in CB buffer [10 mM Mes (pH 6.2), 140 mM NaCl, 2.5 mM EGTA, 5 mM MgCl2] containing 11% sucrose, 3.7% paraformaldehyde, 0.5% glutaraldehyde, and 0.25% Triton. Fixation was stopped with 0.5 mg/mL sodium borohydride in CB for 8 min, followed by washing with PBS and a one hour blocking step with 2% BSA in PBS. Fixed cells were incubated with 1:500 dilution of DM1A anti- -tubulin monoclonal antibody (Sigma) in PBS with 2% BSA and washed three times in PBS. QD625-labeled goat F(ab)2, anti-mouse IgG antibodies (HL) (Invitrogen) was diluted 1:400 in PBS with 6% BSA and applied to cells for one hour. Cells were then dehydrated by sequential washing in for several seconds in 30%, 70%, 90%, and 100% ethanol. Finally, cells were spin coated (500 RPM) with 1 mg/mL PVA.

**Microscope setup.** A custom-built setup around a commercial optical microscope (Zeiss Axiovert 135) is used to image fluorescent sample in a confocal mode. The sample is illuminated with a $473\ nm$ pulsed picosecond laser diode (Edinburgh Instruments) with a $20\ MHz$ repetition rate and a power of $3\ \mu W$, coupled to a single-mode fiber and focused through a 1.4 numerical aperture objective lens (Plan Apo Vc 100, Nikon) into a stationary PSF. A two-axis piezo stage (P-542.2SL XY, Physik Instrumente) scans the sample row-by-row as it is illuminated. The fluorescence is collected by the same objective lens and filtered with a dichroic mirrors and filters (FF509-FDi01, SP01-785RS, BLP01-532R, Semrock). A Galilean beam expander (BE05-10-A, Thorlabs) is placed following the relay lens to magnify the imaged fluorescence spot on to a fiber bundle (A.R.T. Photonics GmbH, Germany). This fiber bundle consists of multimode $100/110\ \mu m$ core/clad fibers, fused together on the entrance side. On the exit side the fibers fan-out to individual multimode fibers; 14 of those guide light from the image plane to individual single-photon avalanche photodiodes (SPCM-AQ4C, Perkin-Elemer). For a detailed characterization of the fiber bundle setup see the Supplementary Information in reference [12] .

**Data acquisition and analysis.** A time-correlated single-photon counting board is used for data acquisition in absolute timing mode (DPC-230, Becker & Hickl GmbH). An excitation pulse trigger is synchronized and recorded at every 40th pulse ($0.5\ MHz$). The correlation analysis, ISM, Q-ISM and Fourier reweighting were implemented in a MATLAB script, post-processing the acquired data. Further details about analysis are given in Supplementary Section 2-4.

# References

1. Abbe, E. Beiträge zur Theorie des Mikroskops und der mikroskopischen Wahrnehmung. *Arch. für Mikroskopische Anat.* **9,** 413–418 (1873).

2. Hell, S. W. & Wichmann, J. Breaking the diffraction resolution limit by stimulated emission: stimulated-emission-depletion fluorescence microscopy. *Opt. Lett.* **19,** 780 (1994).

3. Gustafsson, M. G. L. Surpassing the lateral resolution limit by a factor of two using structured illumination microscopy. SHORT COMMUNICATION. *J. Microsc.* **198,** 82–87 (2000).

4. Rust, M. J., Bates, M. & Zhuang, X. Sub-diffraction-limit imaging by stochastic optical reconstruction microscopy (STORM). *Nat. Methods* **3,** 793–5 (2006).

5. Betzig, E. *et al.* Imaging intracellular fluorescent proteins at nanometer resolution. *Science* **313,** 1642–5 (2006).

6. Dertinger, T., Colyer, R., Iyer, G., Weiss, S. & Enderlein, J. Fast, background-free, 3D super-resolution optical fluctuation imaging (SOFI). *Proc. Natl. Acad. Sci.* **106,** 22287–22292 (2009).

7. Schwartz, O. *et al.* Superresolution microscopy with quantum emitters. *Nano Lett.* **13,** 5832–6 (2013).

8. Gatto Monticone, D. *et al.* Beating the Abbe Diffraction Limit in Confocal Microscopy via Nonclassical Photon Statistics. *Phys. Rev. Lett.* **113,** 143602 (2014).

9. Tsang, M. Quantum limits to optical point-source localization. *Optica* **2,** 646 (2015).

10. Tsang, M., Nair, R. & Lu, X.-M. Quantum Theory of Superresolution for Two Incoherent Optical Point Sources. *Phys. Rev. X* **6,** 031033 (2016).

11. Classen, A., von Zanthier, J., Scully, M. O. & Agarwal, G. S. Superresolution via structured illumination quantum correlation microscopy. *Optica* **4,** 580 (2017).

12. Israel, Y., Tenne, R., Oron, D. & Silberberg, Y. Quantum correlation enhanced super-resolution localization microscopy enabled by a fibre bundle camera. *Nat. Commun.* **8,** 1–5 (2017).

13. Aßmann, M. Quantum-Optically Enhanced STORM (QUEST) for Multi-Emitter Localization. *Sci. Rep.* **8,** 7829 (2018).

14. Cui, J.-M., Sun, F.-W., Chen, X.-D., Gong, Z.-J. & Guo, G.-C. Quantum Statistical Imaging of Particles without Restriction of the Diffraction Limit. *Phys. Rev. Lett.* **110,** 153901 (2013).

15. Sheppard, C. J. R. Superresolution in confocal Imaging. *Optik (Stuttg).* **80,** 53–54 (1988).

16. Müller, C. B. & Enderlein, J. Image Scanning Microscopy. *Phys. Rev. Lett.* **104,** 198101 (2010).

17. Rego, E. H. *et al.* Nonlinear structured-illumination microscopy with a photoswitchable protein reveals cellular structures at 50-nm resolution. *Proc. Natl. Acad. Sci. U. S. A.* **109,**

E135-43 (2012).

18. Blom, H. & Widengren, J. Stimulated Emission Depletion Microscopy. *Chem. Rev.* **117,** 7377–7427 (2017).

19. Sheppard, C. J. R., Mehta, S. B. & Heintzmann, R. Superresolution by image scanning microscopy using pixel reassignment. *Opt. Lett.* **38,** 2889 (2013).

20. Gu, M. & Sheppard, C. J. R. Confocal fluorescent microscopy with a finite-sized circular detector. *J. Opt. Soc. Am. A* **9,** 151 (1992).

21. Ward, E. N. & Pal, R. Image scanning microscopy: an overview. *J. Microsc.* **266,** 221–228 (2017).

22. York, A. G. *et al.* Instant super-resolution imaging in live cells and embryos via analog image processing. *Nat. Methods* **10,** 1122–1126 (2013).

23. Roth, S., Sheppard, C. J., Wicker, K. & Heintzmann, R. Optical photon reassignment microscopy (OPRA). *Opt. Nanoscopy* **2,** 5 (2013).

24. De Luca, G. M. R. *et al.* Re-scan confocal microscopy: scanning twice for better resolution. *Biomed. Opt. Express* **4,** 2644–56 (2013).

25. Laporte, G. P. J., Stasio, N., Sheppard, C. J. R. & Psaltis, D. Resolution enhancement in nonlinear scanning microscopy through post-detection digital computation. *Optica* **1,** 455 (2014).

26. Gregor, I. *et al.* Rapid nonlinear image scanning microscopy. *Nat. Methods* **14,** 1087–1089 (2017).

27. Kimble, H. J., Dagenais, M. & Mandel, L. Photon Antibunching in Resonance Fluorescence. *Phys. Rev. Lett.* **39,** 691–695 (1977).

28. Basche ', T., Moerner, W. E., Orrit, M. & Talon, H. Photon Antihunching in the Fluorescence of a Single Dye Molecule Trapped in a Solid. **69,** (1992).

29. Michler, P. *et al.* A quantum dot single-photon turnstile device. *Science* **290,** 2282–5 (2000).

30. Brouri, R., Beveratos, A., Poizat, J.-P. & Grangier, P. Photon antibunching in the fluorescence of individual color centers in diamond. *Opt. Lett.* **25,** 1294 (2000).

31. Grußmayer, K. S. & Herten, D.-P. Time-resolved molecule counting by photon statistics across the visible spectrum. *Phys. Chem. Chem. Phys.* **19,** 8962–8969 (2017).

32. Portaluppi, D., Conca, E. & Villa, F. 32 × 32 CMOS SPAD Imager for Gated Imaging, Photon Timing, and Photon Coincidence. *IEEE J. Sel. Top. Quantum Electron.* **24,** 1–6 (2018).

33. Castello, M. *et al.* Image Scanning Microscopy with Single-Photon Detector Array. *bioRxiv* 335596 (2018). doi:10.1101/335596

34. Antolovic, I. M., Burri, S., Bruschini, C., Hoebe, R. A. & Charbon, E. SPAD imagers for super resolution localization microscopy enable analysis of fast fluorophore blinking. *Sci. Rep.* **7,** 44108 (2017).

35. York, A. G. *et al.* Resolution doubling in live, multicellular organisms via multifocal structured illumination microscopy. *Nat. Methods* **9,** 749–754 (2012).

36. Schwartz, O. *et al.* Colloidal quantum dots as saturable fluorophores. *ACS Nano* **6,** (2012).

37. Sýkora, J. *et al.* Exploring Fluorescence Antibunching in Solution To Determine the Stoichiometry of Molecular Complexes. (2007). doi:10.1021/AC062024F

38. Eisaman, M. D., Fan, J., Migdall, A. & Polyakov, S. V. Invited Review Article: Single-photon sources and detectors. *Rev. Sci. Instrum.* **82,** 071101 (2011).

39. Hänninen, P. E., Schrader, M., Soini, E. & Hell, S. W. Two-photon excitation fluorescence microscopy using a semiconductor laser. *Bioimaging* **3,** 70–75 (1995).

40. Hell, S. W., Soukka, J. & Hänninen, P. E. Two- and multiphoton detection as an imaging mode and means of increasing the resolution in far-field light microscopy: A study based on photon-optics. *Bioimaging* **3,** 64–69 (1995).

41. Mertz, J. *Introduction to optical microscopy*. (Roberts, 2010).

42. Gonzalez, Rafael, C. *Digital image processing*. (Prentice hall, 2016).

43. Donnert, G., Eggeling, C. & Hell, S. W. Major signal increase in fluorescence microscopy through dark-state relaxation. *Nat. Methods* **4,** 81–86 (2007).

## Acknowledgments

The authors would like to thank Dr. Yuval Ebenstein for the preparation of biological samples and Dr. Stella Izhakov for synthesizing the QDs used in this work. This work was supported by the ERC consolidator grant ColloQuanto, ERC grant QUAMI, the ERC-POC project "SFICAM", the ICore program of the ISF, the Crown Photonics Center and the Israeli ministry of science. R.L. and A.K.P. gratefully acknowledge the hospitality of the Weizmann Institute of Science and the support of National Science Centre (Poland) Grants No. 2015/17/D/ST2/03471 and No. 2015/16/S/ST2/00424, the Polish Ministry of Science and Higher Education, and the FIRST TEAM program of the Foundation for Polish Science co-financed by the European Union under the European Regional Development Fund.

## Contributions

R.T., Y.I., Y.S. and D.O. proposed and designed the experiment. R.T., U.R., B.R., Y.I. and R.L. performed the experimental work. R.T., U.R., B.R., A.K. and R.L. performed the data analysis. R.T. wrote the manuscript with significant contributions from all authors.

# Super-resolution enhancement by quantum image scanning microscopy

Ron Tenne[1*], Uri Rossman[1*], Batel Rephael[1*], Yonatan Israel[1,2], Alexander Krupinski-Ptaszek[3], Radek Lapkiewicz[3], Yaron Silberberg[1], Dan Oron[1]

[1] - Department of Physics of Complex Systems, Weizmann Institute of Science, Rehovot 76100, Israel

[2] - Department of Physics, Stanford University, Stanford, CA 94305, USA

[3] - Institute of Experimental Physics, Faculty of Physics, University of Warsaw, Pasteura 5, 02-093 Warsaw, Poland

[*] - These authors contributed equally to this work

Correspondence and requests for materials should be addressed to D.O. (email: dan.oron@weizmann.ac.il)

Supplementary section 1 – Theory of Q-ISM

In this note we detail the derivation of analytical expressions for the ISM and Q-ISM images that appear in the main text Eq. 1-4.

Let us consider the case of $M$ identical emitters that lie within a single plane and whose positions are given by the 2D vectors $x_i$ ($i = 1..M$). This sample is imaged by a unity magnification imaging system (without inversion) whose incoherent point spread function (PSF) $h(x)$ is symmetric for inversion. Applying a uniform illumination field yields an intensity image

$$G^{(1)}(x) \propto \sum_{i=1}^{M} h(x - x_i). \tag{S1}$$

This is, in fact, a convolution of the sample density, $\rho(x)$, and the system's PSF so that

$$G^{(1)}(x) \propto [\rho * h](x), \tag{S2}$$

where $\rho(x) = \sum_{i=1}^{M} \delta(x - x_i)$ and $*$ stands for the convolution operator.

In standard image scanning microscopy (ISM), presented schematically in figure 1a of the main text, the sample is illuminated with a focused laser beam whose intensity in the sample plane is given by $h_{exc}(x)$ centered around $x = 0$.

As the sample position is scanned to $x_s$ (axes in opposite direction of $x_i$) the position of the emitters shift to $x_i - x_s$. The photoluminescence signal is recorded by an array of point-like detectors, whose image plane positions are denoted by $\bar{x}_\alpha$ ($\alpha = 1,..,K$). The measured intensity in detector $\alpha$ for a scan position $x_s$, $G_\alpha^{(1)}(x_s)$, is given by

$$G_\alpha^{(1)}(x_s) \propto \sum_{i=1}^{M} h_{exc}(x_i - x_s) \cdot h(\bar{x}_\alpha - (x_i - x_s)). \tag{S3}$$

The first term in the product appearing in the right hand side of Eq. S3 is proportional to the probability of an emitter, currently in position $x_i - x_s$ to absorb a photon whereas the second term is proportional to the probability for a photon emitted from the same position to be detected by a detector at $\bar{x}_\alpha$.

Although it is not crucial for the operation of ISM, we simplify the following expressions by assuming $h_{exc} \sim h$. This approximation is justified in cases where the Stokes shift is small and both the collection and excitation are performed with the same objective lens. Under these conditions and using the inversion symmetry of $h$, Eq. S3 turns into

$$G_\alpha^{(1)}(x_s) \propto \sum_{i=1}^{M} h_\alpha(x_s - x_i), \tag{S4}$$

where we have defined the per-detector ISM PSF for each detector, $h_\alpha(x) \equiv h(x) \cdot h(\bar{x}_\alpha - x)$. From symmetry consideration we can deduce that the effective PSFs are centered around $\frac{\bar{x}_\alpha}{2}$ and for a practical microscope will be narrower than the original PSF. In cases where the original system PSF can be well approximated with a Gaussian function, $h_\alpha$ is narrower than $h$ by a factor $\sqrt{2}$.

An ISM image is formed by centering each of the $K$ images and summing them together

$$G_{ISM}^{(1)}(x_s) \propto \sum_{\alpha=1}^{K} G_\alpha^{(1)}\left(x_s - \frac{\bar{x}_\alpha}{2}\right) \qquad \text{(S5)}$$
$$= \sum_{i=1}^{M} \sum_{\alpha=1}^{K} h\left(x_i - x_s - \frac{\bar{x}_\alpha}{2}\right) \cdot h\left(x_i - x_s + \frac{\bar{x}_\alpha}{2}\right).$$

The right-hand side of Eq. S5 can be expressed as a convolution in

$$G_{ISM}^{(1)}(x_s) \propto [\rho * h_{ISM}](x_s) \qquad \text{(S6)}$$

where

$$h_{ISM}(x) = \sum_{\alpha=1}^{K} h\left(x - \frac{\bar{x}_\alpha}{2}\right) \cdot h\left(x + \frac{\bar{x}_\alpha}{2}\right) = \sum_{\alpha=1}^{K} h_\alpha\left(x - \frac{\bar{x}_\alpha}{2}\right) \qquad \text{(S7)}$$

is the effective PSF for ISM. Since this PSF is a sum of the per-detector effective ISM PSFs, centered around $x = 0$, it retains roughly the same resolution enhancement in the image generated by each detector with a highly increased signal level.

If the emitters labeling the sample are single photon emitters, they emit, at most, a single photon when excited with a single laser pulse shorter than the excitation lifetime. To observe photon correlation, our analysis focuses on the probability of detecting simultaneous photon pairs, defined as photon pairs detected during the time between the same excitation pulses of a pulsed laser. The probability for detecting such a simultaneous photon pair with two different detectors $\alpha$ and $\beta$ is

$$G_{\alpha\beta}^{(2)}(x_s) \propto \sum_{\substack{i \neq j}}^{M} h(x_i - x_s) \cdot h\big(\bar{x}_\alpha - (x_i - x_s)\big) \cdot h(x_j - x_s) \qquad \text{(S8)}$$
$$\cdot h\big(\bar{x}_\beta - (x_j - x_s)\big),$$

where the summation is over all unequal index pairs $i, j$ of emitters. The absence of equal index terms $(i = j)$ in the sum is the manifestation of antibunching; it is impossible for an emitter to emit two photons after a single short laser pulse.

For each of the $\frac{K \cdot (K-1)}{2}$ detector pairs we can extract the number of missing photon pair by subtracting $G_{\alpha\beta}^{(2)}(x_s)$ from the probability for a photon pair under Poissonian photon statistics

$$\Delta G^{(2)}_{\alpha\beta}(x_s) = G^{(1)}_\alpha(x_s) \cdot G^{(1)}_\beta(x_s) - G^{(2)}_{\alpha\beta}(x_s)$$
$$= \sum_{i=1}^{M} [h(x_i - x_s)]^2 \cdot h(\bar{x}_\alpha - (x_i - x_s)) \qquad (S9)$$
$$\cdot h(\bar{x}_\beta - (x_i - x_s)).$$

This is an expression for the number of 'missing' photon pairs for a position scan $x_s$ as observed by the detector pair $\alpha, \beta$. The product of the four shifted PSFs can be written in terms of the per-detector effective ISM PSFs as

$$\Delta G^{(2)}_{\alpha\beta}(x_s) = \sum_{i=1}^{M} h_\alpha(x_s - x_i) \cdot h_\beta(x_s - x_i) \equiv \sum_{i=1}^{M} h^{(2)}_{\alpha\beta}(x_s - x_i). \qquad (S10)$$

where we have defined a per-detector pair Q-ISM PSF as $h^{(2)}_{\alpha\beta}$ for a detector pair $\alpha, \beta$, that is a product of two of the effective ISM PSFs for the corresponding detectors. In the case of a Gaussian system PSF, $h^{(2)}_{\alpha\beta}$ is a factor of $\sqrt{2}$ narrower than $h_\alpha$. Thus, the image formed in this measurement has a two-fold resolution enhancement when compared to the standard microscope image, $G^{(1)}(x)$. The center of $h^{(2)}_{\alpha\beta}$, in the Gaussian case, is the average of centers for the four functions appearing in the right-hand side of Eq. S9, $\frac{\bar{x}_\alpha + \bar{x}_\beta}{4}$.

In order to obtain one super-resolved image with a high signal-to-noise ratio (SNR) the $\frac{K \cdot (K-1)}{2}$ different images should be shifted and summed, as performed in standard ISM:

$$G^{(2)}_{Q-ISM}(x_s) = \sum_{\alpha \neq \beta} \Delta G^{(2)}_{\alpha\beta}(x_s - \bar{x}_{\alpha\beta}) \qquad (S11)$$

where $\bar{x}_{\alpha\beta}$ are the appropriate shift vectors for the image formed by the missing pairs detected in the $\alpha$ and $\beta$ detectors.

## Supplementary section 2 – Fourier reweighting

### The motivation for Fourier reweighting

In a similar manner to ISM, the convolution of the object with the point spread function (PSF) of Q-ISM results in an attenuation of the higher spatial frequencies within the pass band of the PSF. It is therefore useful to perform a Fourier reweighting (FR) procedure to enhance the image resolution, i.e. to reweight the image Fourier representation in order to enhance higher spatial frequencies.

We start by examining the spatial frequency extent of the images measured with Q-ISM in order to estimate the potential for resolution enhancement. To analyze the images in the Fourier domain we use the following definition of a two-dimensional Fourier transform

$$S(\vec{\kappa}) = \mathcal{F}\{I(\vec{r})\} = \int_{-\infty}^{\infty} dx \int_{-\infty}^{\infty} dy \left[ I(\vec{r}) \cdot e^{-2\pi i \vec{\kappa} \cdot \vec{r}} \right], \tag{S12}$$

where $\mathcal{F}\{\}$ denotes the Fourier transform, $\vec{r}$ and $\vec{\kappa}$ are two-dimensional vectors in position and spatial frequency space respectively.

The point spread function (PSF) of a wide-field microscope is limited, in the Fourier domain, by a circle of radius $\kappa_c = \frac{NA}{\lambda}$, where $NA$ is the numerical aperture of the imaging system and $\lambda$ is the wavelength of the light. Therefore, imaging a coherent signal can be considered as a low pass filter of the object in the spatial frequency space. When imaging an incoherent source of light the real space PSF takes the form[41]

$$h_1(r) \propto \frac{J_1^2(2\pi \kappa_c r)}{r^2}, \tag{S13}$$

where $J_1$ is the first order Bessel function of the first kind and $r = |\vec{r}|$. The Fourier transform of this PSF is termed the optical transfer function $OTF_1(\kappa)$. Its absolute value, the modulation transfer function, $MTF_1(\kappa)$, is presented in figure **S1**a. It decays gradually with higher frequencies and reaches a zero value for all frequencies outside of the pass band defined by $|\vec{\kappa}| \leq 2\kappa_c$.

In a confocal setup with, for simplicity, one ideal point detector located on the optical axis, the ISM PSF is expressed in real space as

$$h_\alpha^{(1)}(r) \propto [h_1(r)]^2 \tag{S14}$$

and in the Fourier space as

$$OTF_\alpha^{(1)}(\vec{\kappa}) = \mathcal{F}\left\{h_\alpha^{(1)}(r)\right\} \propto h_1(\vec{\kappa}) * h_1(\vec{\kappa}), \tag{S15}$$

where $*$ stands for the convolution operator. According to the definition of the convolution operator the highest frequency with a non-zero value, defining the OTF pass band in the case of ISM, is larger by a factor of 2 compared to the wide-field case, $|\vec{\kappa}| \leq 4\kappa_c$.

If the above-mentioned point detector is capable of measuring antibunching it allows us to construct a Q-ISM image whose PSF is given by

$$h^{(2)}_{\alpha\alpha}(r) \propto [h_1(r)]^4. \tag{S16}$$

In Fourier space, the Q-ISM can be expressed as

$$OTF^{(2)}_{\alpha\alpha}(\vec{\kappa}) = \mathcal{F}\left\{h^{(2)}_{\alpha\alpha}(r)\right\} \propto h^{(1)}_{\alpha}(\vec{\kappa}) * h^{(1)}_{\alpha}(\vec{\kappa}), \tag{S17}$$

extending to spatial frequencies up to $8\kappa_c$, a radius 4 time larger than for the widefield microscope. The absolute value of $OTF^{(2)}_{\alpha\alpha}(\vec{\kappa})$, $MTF^{(2)}_{\alpha\alpha}(\vec{\kappa})$, is shown in figure **S1**b with the experimental $NA$ estimated for our system, 0.83. Cross-sections of the MTF for wide-field, ISM and Q-ISM imaging are shown in figure **S1**c.

An extension of the pass band by a factor of 4 suggests an up to 4-fold enhancement of the resolution is possible in Q-ISM. However, since the MTF for Q-ISM decays in a more rapid manner than that of the wide-field MTF (scaled to the extent of the pass band) the resolution enhancement in Q-ISM is lower than 4 and is expected to be roughly 2 even with an ideal point detector. We can gain some quantitative impression noting that the MTF reaches 1% of its peak value at $k = 3\kappa_c$ and $k = 4.26\kappa_c$ for the ISM and Q-ISM methods respectively.

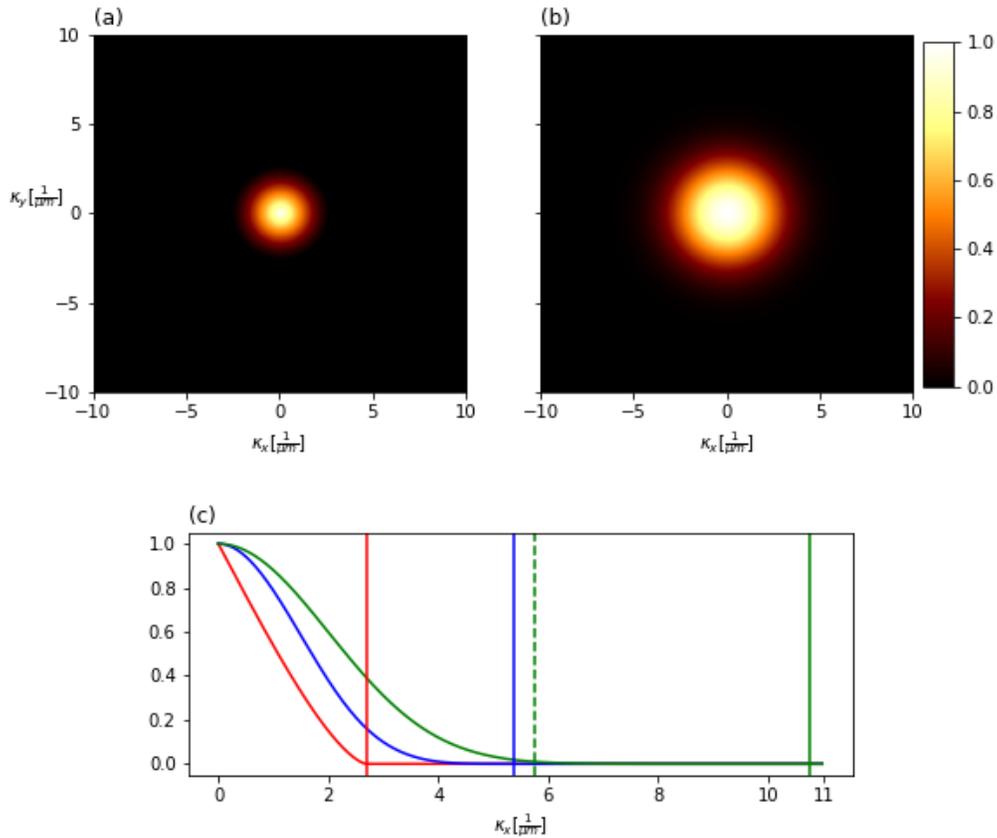

**Figure S1|** **(a)** $MTF$ of a wide-field microscope ($NA = 0.83$) **(b)** $MTF$ of Q-ISM for the same $NA$ as in (a). **(c)** Radial cross sections of the $MTF$s for wide-field microscopy, ISM and Q-ISM are shown in red, blue and green lines respectively. The pass band radius for

each case is represented by a vertical line with a matching color. The dashed vertical line corresponds to a 1% of the peak value $\kappa = 4.26 \cdot \kappa_c \simeq 5.73 \frac{1}{\mu m}$.

## The Wiener filter and its variations

Assuming an ideal, noiseless, imaging system and exact knowledge of the system's $OTF$, one could simply divide the Fourier transform of the image by the $MTF$ to retrieve the a Fourier limited super-resolved image. In this case, termed inverse filtering, the Fourier transform of the image $S(\kappa)$ is multiplied by a Fourier reweighting function $W_{inv}(\kappa) = \frac{1}{OTF(\kappa)}$. However, at high spatial frequencies, strongly amplified by $W_{inv}$, the signal-to-noise (SNR) ratio is usually low. As a result, in a practical imaging scenario this direct deconvolution procedure greatly amplifies the noise and therefore fails to produce a high SNR image. A more practical approach is to apply a Wiener filter on the image with a smoothly decreasing amplification factor at higher frequencies, according to the SNR[42]. A Wiener filter is an optimal solution for an image whose noise is uncorrelated with the signal, such as camera readout noise, an additive Gaussian noise. However, in Q-ISM the dominant noise source is Poissonian shot noise in the detection of photon pairs (see also Supplementary section 6). As a result the noise is highly correlated to the signal level.

We therefore employ here a modified version of the Wiener filter in which the reweighting function is

$$W(\kappa) = \begin{cases} \dfrac{1}{MTF_{Q-ISM}(\vec{\kappa}) + \epsilon \cdot \dfrac{|\vec{\kappa}|}{\kappa_{max}}} & |\vec{\kappa}| \leq \kappa_{max} \\ 0 & \text{otherwise} \end{cases}, \qquad (S18)$$

where $MTF_{Q-ISM}(\vec{\kappa})$ is an estimate of the Q-ISM MTF[26]. Both $\epsilon$ and $\kappa_{max}$ are parameters that should be optimized for the typical SNR of the method. In this work we chose to work with parameter values $\epsilon = 0.3$ and $\kappa_{max} = 8 \frac{1}{\mu m}$ for all images, including those shown in figure 2 and figure 3 of the main text. Note that the value $\kappa_{max}$ is approximately $6 \cdot \kappa_{max}$ for our system.

Another important component in evaluating $W(\kappa)$ is estimating the MTF. For a robust operation of Fourier reweighting we chose to use calibrated data sets measured in the same measurement day to estimate $MTF_{Q-ISM}(\vec{\kappa})$. Figure **S2** presents the Q-ISM $PSF$ (calibration data, as described in Supplementary Section 4), $MTF_{Q-ISM}(\vec{\kappa})$ and $W(\kappa)$ used for the Fourier reweighting procedure performed for figure 2 of the main text.

We note that the approach used here for Fourier reweighting and pixel reassignment may not be the optimal one to construct a low SNR high-resolution image from the multiple photon correlation images. In this work, our focus is to demonstrate the experimental method of Q-ISM. Therefore, we did not make a comparative study of post-processing techniques to find an optimal approach.

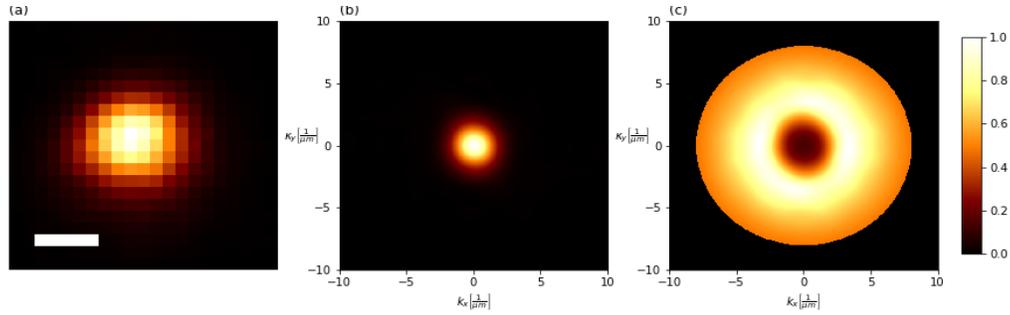

**Figure S2| Constructing a Fourier reweighting filter.** A variant of a Wiener filter is constructed using a calibration measurement for figure 2 of the main text. **(a)** Mean photon pair calibration image of an isolated fluorescent bead (F8786, Thermo Fisher). Further details about the calibration process can be found in Supplementary Section 4. **(b)** Corresponding $MTF_{Q-ISM}(\vec{\kappa})$ **(c)** Reweighting function for $\epsilon = 0.3$ and $\kappa_{max} = 8 \frac{1}{\mu m}$. Scale bar in (a): $0.25\ \mu m$.

### A Fourier reweighting protocol for Q-ISM images

In a similar manner to the procedure used by Gregor et al.[26] Fourier reweighting for Q-ISM was done as follows:

1. Zero-pad the calibration data (PSF) and the image to oversample in Fourier space.

2. Compute $S(\kappa) = \mathcal{F}\{I(r)\}$

3. Compute $|S(\kappa)|$ and $\text{Arg}(S(\kappa))$ to reweight only the modulus.

4. Construct the reweighting function $W(\kappa)$:

    a. Fourier transform the mean photon pair calibration image from the same measurement day and take the modulus $MTF_{Q-ISM}(\vec{\kappa}) = |\mathcal{F}\{PSF(r)\}|$

    b. Compute $W(\kappa)$ according to Eq.S18

5. Retrieve the original image as $I_{FR} = |\mathcal{F}^{-1}\{W(\kappa) \cdot |S(\kappa)| \cdot \exp(i\text{Arg}(S(\kappa)))\}|$

We note that although the choice of parameter values for $\kappa_{max}$ and $\epsilon$ affects the reweighted image appearance, this dependence is smooth. The resolution of the image improves with increasing $\kappa_{max}$ up to high values where an additional increase mostly adds to the image noise. The practical choice of $\epsilon$ depends on the SNR and may depend on experimental parameters such illumination power, quantum yield of fluorescent markers and pixel dwell time.

Further experimental results of FR Q-ISM images, shown in figure S3, present a clear improvement in the resolution. Alongside this resolution enhancement appears a clear ringing artifact in the background which may interfere with the observation of dim objects. We note that this phenomenon is not particular to Q-ISM but is a general issue in Fourier reweighting. A Fourier reweighted PSF carries less intensity in the narrower central peak while the intensity of the outer rings grows.

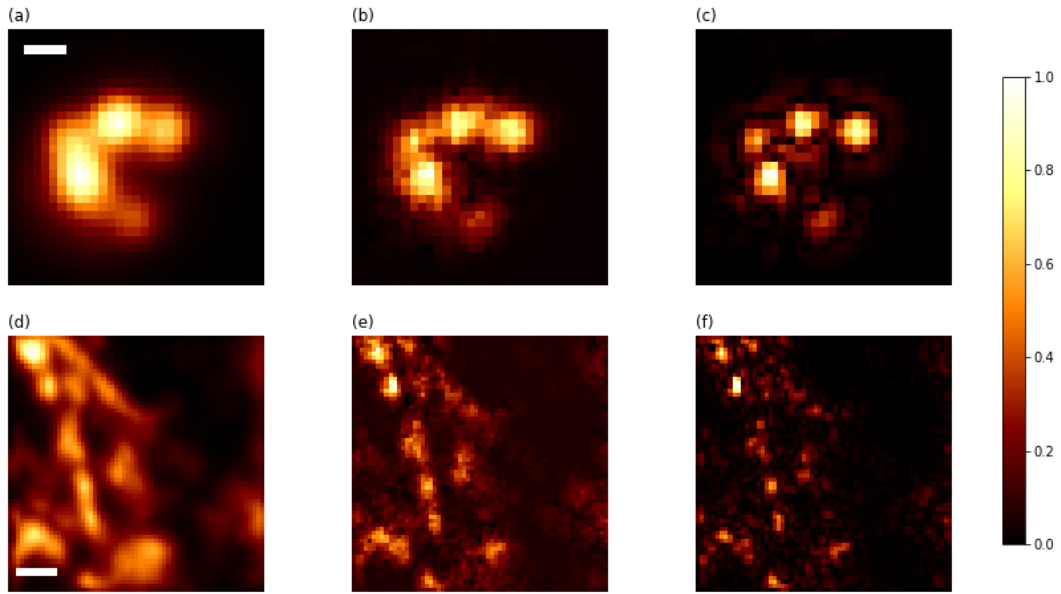

**Figure S3|** Comparison of ISM (left), Q-ISM (center) and FR Q-ISM (right) images for a few QDs (a)-(c) and micro-tubules labeled with QDs (d)-(f. (a), (d) ISM images (b), (e) Q-ISM images (c), (f) FR Q-ISM images, $\epsilon = 0.3$, $\kappa_{max} = 8 \frac{1}{\mu m}$. Scale bars: $0.25 \mu m$ (a)-(c), $0.5 \mu m$ (d)-(f).

## Supplementary section 3 – Data analysis for ISM and Q-ISM

During a sample scan, the single photon avalanche detectors (SPADs) continuously collect photons whose arrival times are measured by the time correlated single photon counting (TCSPC) card and recorded onto a digital file. The TCSPC channels fed by the SPAD output are referred to, in the following, as SPAD channels.

Two additional channels in the 16 channels TCSPC card (DPC230, Becker & Hickl) are used in order to sync the measurement with the laser trigger and the scan stage movement. The laser trigger is diluted by a function generator from a 20 MHz rate pulse train into a 0.5 $MHz$, 4V amplitude pulse train and input into one of the TCSPC channels. The trigger is diluted in order not to saturate the TCSPC card and avoid large data files. Another channel in the TCSPC card times a trigger pulse train synchronized to the sample scanning stage moves. In this manner we measure the stage movement, pulse trigger and photon detection timing with the same clock, that of the TCSPC card.

### ISM image construction

In post processing performed by a MATLAB script, the data is parsed according to sample scan trigger timing. To create the confocal photoluminescence (PL) image for each detector, we count the detections in each of the 14 SPAD channels per scan move. Ordering the number of detected photons according to the actual stage position yields a single detector confocal scan image signified by $G_\alpha^{(1)}(x_s)$ in Supplementary Section 1. To generate an ISM image, we shift each of these images according to the pre-calibrated translation vector, $\overrightarrow{\delta r_\alpha}$, (Supplementary Section 4) and sum them together to obtain

$$G_{ISM}^{(1)}(x_s) = \sum_{\alpha=1}^{14} G_\alpha^{(1)}(x_s - \overrightarrow{\delta r_\alpha}). \tag{S19}$$

Since the translation vectors, in general, contain fractional pixel values, the translated sub-image values are linearly interpolated to the original grid.

### Q-ISM image construction

The process of generating a Q-ISM image begins with the calculation of the second order correlation function for each of the detector pairs in each of the scan steps, $G_{\alpha\beta}^{(2)}(\tau, x_s)$. We assign each photon detection to its preceding laser trigger, taking into account the fact that the trigger pulse train has been diluted by adding equally spaced 'fake' pulses between two measured triggers. For a detector pair, we histogram the time delay, in units of pulse period ($t_p$), between all photons in the first channel with all the photons in the second channel. To avoid artifacts due to the SPADs' dead time we do not generate photon correlation data for a channel with itself ($G_{\alpha\alpha}^{(2)}$). An antibunching signal is computed by subtracting the number of photon pairs arriving after the same pulse, from an average number of photon pairs with longer delay times

$$\Delta G_{\alpha\beta}^{(2)}(x_s) = \frac{1}{n_2 - n_1 + 1}\left[\sum_{j=n_1}^{n_2} G_{\alpha\beta}^{(2)}(j \cdot t_p, x_s)\right] - G_{\alpha\beta}^{(2)}(0, x_s) , \tag{S20}$$

where in our analysis $n_1 = 4$ and $n_2 = 50$.

The 91, distinct detector pair, sub-images are translated by the pre-calibrated translation vectors $\overrightarrow{\delta r}_{\alpha\beta}$ and summed into a single image

$$G^{(2)}_{Q-ISM}(x_s) = \sum_{\alpha \neq \beta} \Delta G^{(2)}_{\alpha\beta}(x_s - \overrightarrow{\delta r}_{\alpha\beta}) \qquad (S21)$$

## Supplementary section 4 - Calibration process for pixel reassignment

When performing Q-ISM and ISM confocal scans, one needs to merge the numerous images produced by each detector pair and single detectors respectively. In the following we refer to these images as sub-images. In its simplest form, the pixel-reassignment algorithm requires prior knowledge about the 2D translation vectors for each of the sub-images in a scan. This is especially important in the case of a fiber-bundle detector since, unlike a camera, the pixels are not positioned precisely on a lattice. We therefore perform a reference scan measurement of a single isolated fluorescent bead with a diameter of $20 nm$, much smaller than the fluorescence wavelength (F8786, Thermo Fisher).

We can model a small bead as a large number of emitters, $M \gg 1$, all positioned at $x_i = 0$. The expression for the ISM sub-images is calculated by plugging these positions into Eq.S4 in supplementary Section 1,

$$G^{(1)}_{\alpha-Bead}(x_s) = M \cdot h(x_s) \cdot h(\bar{x}_\alpha - x_s). \tag{S22}$$

Since we perform the calibration measurement well below saturating conditions and the number of emitters in a bead is large, the antibunching signal from the bead is negligible. Therefore, to calibrate the Q-ISM translation vectors we construct sub-images of the mean number of simultaneous photon pairs and mathematically show their equivalence to imaging one single-photon emitter. The mean simultaneous photon pairs sub-image of a single bead is given by

$$G^{(2)}_{\alpha\beta-Bead}(x_s) = M \cdot (M-1) \cdot h(x_s) \cdot h(\bar{x}_\alpha - x_s) \cdot h(x_s) \cdot h(\bar{x}_\beta - x_s), \tag{S23}$$

where we have summed Eq.S8 over the same $M$ emitters.

In more precise terms, we would like to show that the sub-images $G^{(1)}_{\alpha-Bead}(x_s)$ and $G^{(2)}_{\alpha\beta-Bead}(x_s)$ are equivalent to the sub-images $G^{(1)}_{\alpha-SP}(x_s)$ and $\Delta G^{(2)}_{\alpha\beta-SP}(x_s)$ respectively, where the last two are calculated for the case of one single photon emitter positioned at $x_1 = 0$. For this purpose we plug $M = 1$ and $x_1 = 0$ in Eq.S4 and Eq.S9 and obtain

$$G^{(1)}_{\alpha-SP}(x_s) = h(x_s) \cdot h(\bar{x}_\alpha - x_s) \tag{S24}$$

and

$$\Delta G^{(2)}_{\alpha\beta-SP}(x_s) = [h(x_s)]^2 \cdot h(\bar{x}_\alpha - x_s) \cdot h(\bar{x}_\beta - x_s), \tag{S25}$$

leading to the proportionality relations

$$G^{(1)}_{\alpha-Bead}(x_s) = M \cdot G^{(1)}_{\alpha-SP}(x_s) \tag{S26}$$

and

$$G^{(2)}_{\alpha\beta-Bead}(x_s) = M \cdot (M-1) \cdot \Delta G^{(2)}_{\alpha\beta-SP}(x_s). \tag{S27}$$

Indeed, we can see that $G^{(1)}_{\alpha-Bead}(x_s)$ and $G^{(2)}_{\alpha\beta-Bead}(x_s)$ are translated precisely as the ISM and Q-ISM sub-images.

To estimate the ISM translation vectors we begin by summing all ISM sub-images without translations and fit those to with a 2D Gaussian to find the center position of the pattern, $\vec{r}_0$ (Fig.**S4**a). We translate all sub-images to this arbitrary position; choosing a different position would only shift the entire resulting image together. The sub-image formed by each detector is also fit with a 2D Gaussian and its center position is marked with $\vec{r}_\alpha$ (red circles in Fig.**S4S4**b-d). Finally, the translation vectors are calculated as

$$\overrightarrow{\delta r_\alpha} = \vec{r}_0 - \vec{r}_\alpha, \tag{S28}$$

where $\alpha$ is an integer ranging between 1 and 14.

We perform a similar procedure for the simultaneous photon pair images: the sum of all non-translated images is fit to a 2D Gaussian function and its center is marked as $\vec{r}_{00}$. The sub-image formed by each detector pair $\alpha,\beta$ is also fit with a 2D Gaussian function and its center is marked as $\vec{r}_{\alpha\beta}$. The translation vectors are calculated as

$$\overrightarrow{\delta r_{\alpha\beta}} = \vec{r}_{00} - \vec{r}_{\alpha\beta}, \tag{S29}$$

where both $\alpha$ and $\beta$ are integers ranging between 1 and 14.

In principle, the translation vectors can be sensitive to the optical alignment of the system and the shape of the fluorescent bead. We have therefore tested the repeatability of this procedure and found that the translation vectors for measurements of 3 different beads agreed up to ~0.5%.

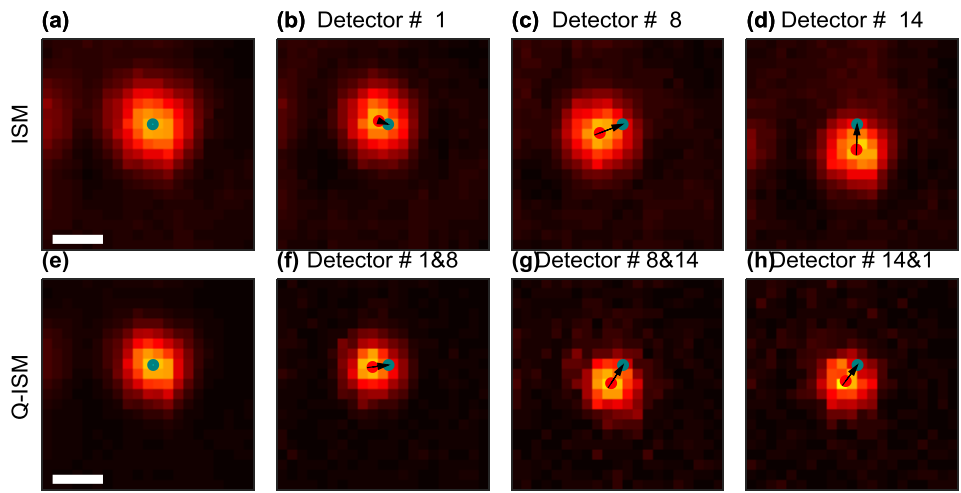

**Fig. S4| PL reassignment process. (a)** A $1\mu m$ by $1\mu m$ confocal scan ($50 nm$ steps, 100ms pixel dwell time) of a single fluorescent bead. The image is a confocal scan image of the sum of all individual detectors. The blue circle represents the center of a 2D Gaussian fit of the images' sum, $\vec{r}_0$. This location is used as a reference position to which all image centers are translated. **(b)-(d)** The confocal scan image of individual detectors numbered 1, 8 and 14 respectively. The red circle in each image represents the center of the image's 2D Gaussian fit. The blue circle as in (a) shows the reference position, $\vec{r}_0$. The black arrows show the translation vectors, $\overrightarrow{\delta r_\alpha}$. **(e)** A summed image of the mean number of photon pairs summed over all detector pairs calculated from the same data used to produce panels (a)-(d). The blue circle shows the center of a 2D Gaussian fit of the image. **(f)-(h)** A scanned image of the mean number of photon pairs for a specific detector pair (numbers noted above each panel). As in (b)-(d) the red circle represents the center of a 2D Gaussian fit and the black arrows correspond to the translation vectors $\overrightarrow{\delta r_{\alpha\beta}}$. Scale bar: $0.25\mu m$.

Supplementary section 5 – Quantitative assessment of resolution enhancement

As discussed in Supplementary Section 4, the mathematical expression for ISM and Q-ISM for one single photon emitter are proportional to the calibration images of an isolated fluorescent bead whose diameter is much smaller than the fluorescence wavelength. Therefore, to quantitatively evaluate the resolution enhancement in Q-ISM, we measured a single particle with a much smaller than the PSF of conventional imaging (Fig.*S5*). For this purpose, 20 $nm$ diameter fluorescent beads (F8786, Thermo Fisher Scientific) were favored over quantum dots due to their bright and stable fluorescence. In addition to the standard confocal scans, we captured a wide-field image measured by placing an EMCCD (Rolera Thunder, Qimaging) behind a beam splitter set before the fiber-bundle.

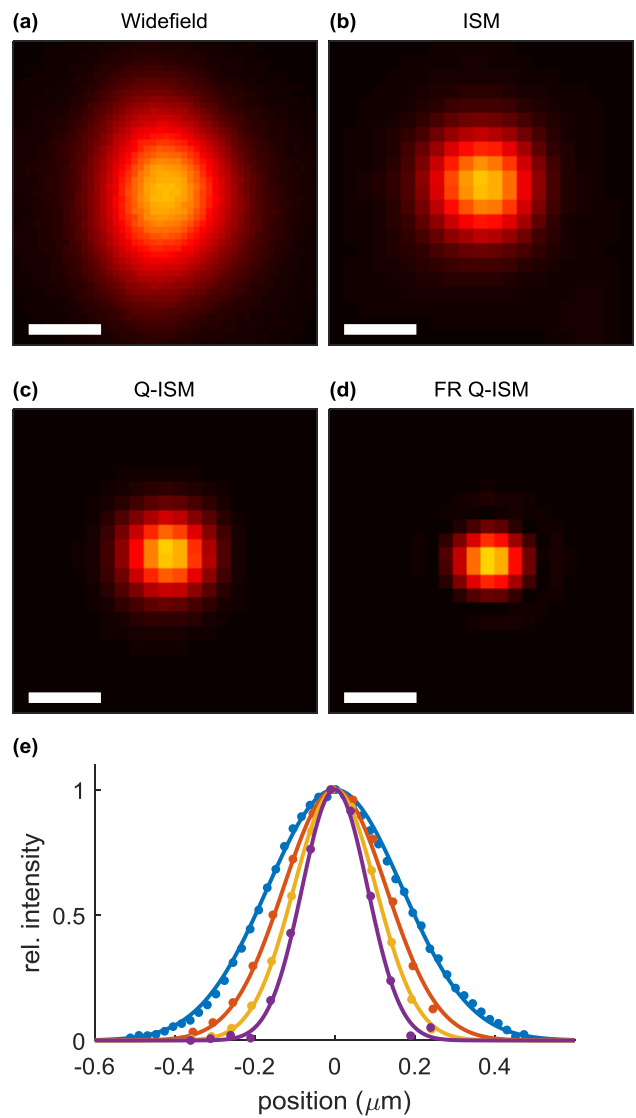

**Fig. S5|** Resolution improvement in Q-ISM. (**a**) An image of a single, effectively point-like, fluorescent bead, captured by an EMCCD camera placed at the imaging plane of the

optical setup. (**b-d**) scan results of the same bead in (a). (**b**) ISM image. (**c**) Q-ISM calibration image (Supplementary Section 4). (**d**) Fourier reweighted Q-ISM calibration image. (**e**) Cross sections of the images in (a)-(d) through the horizontal axis centered at the bead center position. Wide-field image (blue), ISM (red), Q-ISM calibration (yellow) and Fourier reweighted Q-ISM calibration (purple). The solid lines correspond to 1D Gaussian fits to the cross-section data. Scale bars for (a)-(d): 0.25 $\mu m$.

Horizontal cuts from the center of each image in Fig. *S5*a-d are shown in figure *S5*e. Each of the cuts was fit with a 1D Gaussian function to estimate the width of the PSF.

We repeated the measurement for 15 different beads. Average values of $2\sigma$ widths for the wide-field imaging, ISM, Q-ISM and Fourier reweighted (FR) Q-ISM PSFs were $309 \pm 5\ nm$, $240 \pm 15\ nm$, $180 \pm 11\ nm$ and $137 \pm 11\ nm$ respectively. The error given for the PSF widths are the standard deviations of the 15 measurements. The resolution enhancement was, on average, $1.28 \pm 0.07$ from the wide-field image to an ISM image, $1.35 \pm 0.03$ from an ISM image to a Q-ISM image and $1.31 \pm 0.03$ from a Q-ISM image to a Fourier reweighted Q-ISM image. The total resolution enhancement is estimated as $2.27 \pm 0.17$ from the wide-field images to the FR Q-ISM images.

Using the Gaussian fit for the wide-field imaging PSF we estimate the numerical aperture of our imaging system as 0.83, by noting that a Gaussian approximation of an intensity Airy pattern shows $\sigma \cong 0.34 \cdot r_0$, where $r_0$ is the radius of the first dark ring in the Airy pattern. The discrepancy between this value and the upper limit set by the objective lens, 1.4, is probably due to aberrations caused by the magnifying optics.

## Supplementary section 6 – SNR in ISM and Q-ISM

### A discussion of SNR in Q-ISM images

In the following, we derive an analytical expression for the SNR in Q-ISM. The noise in a Q-ISM image is that of the antibunching signal used to construct it. To simplify, we consider $M$ emitters positioned at $x_i = 0$. The photon detection probability for each emitter per pulse is given as $p$. Therefore, the number of photon pairs arriving at some delay longer than a single pulse and the excitation lifetime is

$$G^{(2)}(\infty) = N \cdot (M \cdot p)^2, \tag{S30}$$

where $N$ is the number of pulses on which we integrate, i.e. the pixel dwell time in a scan. Note that this relation and the entire analysis follows we assume that the SPAD detectors are well below saturation.

The number of photon pairs arriving simultaneously, derived from Eq.S8 in Supplementary section 1, is

$$G^{(2)}(0) = N \cdot M(M-1) \cdot p^2, \tag{S31}$$

where we have included the effects of photon antibunching.

The antibunching signal, which serves as the contrast of Q-ISM images, as measured by the dip in the second order correlation function (see Fig.1b of the main text) is

$$\Delta G^{(2)} = G^{(2)}(\infty) - G^{(2)}(0) = N \cdot M \cdot p^2 \tag{S32}$$

The two quantities $G^{(2)}(\infty)$ and $G^{(2)}(0)$ vary stochastically in uncorrelated manner and therefore the variance of $\Delta G^{(2)}$ is

$$V[\Delta G^{(2)}] = V[G^{(2)}(\infty)] + V[G^{(2)}(0)]. \tag{S33}$$

In addition, since $G^{(2)}(\infty)$ can be deduced by averaging over many delays (see Fig. 1b of the main text and supplementary Section 3) its variance can be neglected yielding

$$V[\Delta G^{(2)}] \cong V[G^{(2)}(0)]. \tag{S34}$$

Since the PL signal is at least an order of magnitude larger than the sum of SPAD dark current over the entire array, the only considerable noise source is that of shot noise in the number of measured photon pairs. That is the variance of $\Delta G^{(2)}$ is given by

$$V[\Delta G^{(2)}] \cong V[G^{(2)}(0)] = G^{(2)}(0) = N \cdot M(M-1) \cdot p^2 \cong N \cdot M^2 \cdot p^2, \tag{S35}$$

where in the last approximation we have assumed $M \gg 1$.

Using the last result and Eq.S32, the SNR is estimated as

$$SNR_{Q-ISM} = \frac{\Delta G^{(2)}}{\sqrt{V[\Delta G^{(2)}]}} = \frac{N \cdot M \cdot p^2}{\sqrt{N \cdot M^2 \cdot p^2}} = \sqrt{N} \cdot p. \tag{S36}$$

An interesting consequence of this result is that the SNR is independent of the number of emitters, $M$. This can be explained by considering that for $M \gg 1$ the antibunching signal is a small difference between two large numbers, the number of measured photon pairs. The number of photon pairs scales as the number of emitter pairs squared, $M^2$, representing all the different options for an emission of a photon pair. On the other hand, the difference, the number of missing photon pairs, increases only linearly with $M$ since these are missing photon pair events only from the same emitter. As a result both the missing photon pair signal and the error on the total number of detected photon pairs scale linearly with $M$ and their ratio is independent of $M$.

The SNR does depend critically on $p$, the number of detected photons per emitter per pulse. This probability is a product of the probability of an emitter to absorb a photon, its probability to emit a photon and the probability of collecting and detecting that light. The first term, the absorption probability, is limited by the saturation of the emitter, that is, with a high enough laser power we can excite the emitter once per pulse. However, working close to saturation conditions has a negative impact on the stability and blinking of dye molecules and quantum dots (QDs)[36,43]. We therefore prefer to work at a laser power 2-5 times smaller than the saturation power.

The second term, the probability to emit given that the fluorophore was excited, termed quantum yield (QY), depends on the type of emitters. The QY of dyes, commercially used in super-resolution microscopy, and QDs is usually quite high (>50%). One can consider the emitters 'blinking' by reduction in the number of excitation pulses to those in which the emitter was in the bright state, roughly $\frac{N}{2}$. The third term, the collection and detection efficiency, is usually limited by the finite numerical aperture of the objective lens and detection probability of the SPADs and can be estimated to be between 10% and 20%[12]. Considering all these factors, for a third of the saturation excitation intensity, in the case of high QY markers, $p \sim 0.03$ and can prove challenging to improve.

The potential of improvement with increasing the number of pulses is also limited. The number of pulses per step is a product of the pixel dwell time with repetition rate of laser pulses. The dwell time has to be limited in order to avoid very long detection times and drifts in the sample position. Although the laser repetition rate can be increased and a CW laser can be used, the excitation lifetime limits the amount of photons emitted from a fluorophore and therefore also the number of missing photon pairs.

### A discussion of SNR in higher order correlations images

To expand the discussion, we evaluate the SNR of Q-ISM for correlation orders higher than 2. Using a correlation order $R$ allows to extend the resolution enhancement to $2R$. However, the following estimate of SNR shows that for any order higher than 2 the SNR decreases with a higher density of markers. It may therefore be difficult to extend Q-ISM beyond the second order correlation presented in this work.

The signal of an $R$-order antibunching is the missing events in which $R$ detected photons originate from the same emitter. Therefore, without defining precisely the derivation of this contrast from correlation functions up to $R$ order we estimate the frequency of this events as

$$\Delta G^{(R)} \propto N \cdot M \cdot p^R. \tag{S37}$$

In a similar way to obtaining the $\Delta G^{(2)}$ contrast, previously described in this section, an estimate of $\Delta G^{(R)}$ requires a subtraction of a few terms larger than $\Delta G^{(R)}$; in the order of the number of simultaneous $R$ photon detections events, $N \cdot M^R \cdot p^R$. Therefore considering only shot noise the variance of the $R$ order correlation contrast is

$$V[\Delta G^{(R)}] \propto N \cdot M^R \cdot p^R \qquad (S38)$$

and the SNR is

$$SNR = \frac{\Delta G^{(R)}}{\sqrt{V[\Delta G^{(R)}]}} \propto N^{\frac{1}{2}} \cdot M^{1-\frac{R}{2}} \cdot p^{\frac{R}{2}}. \qquad (S39)$$

Since $p < 1$, the term $p^{\frac{R}{2}}$ becomes exponentially smaller with an increasing correlation order necessitating a longer exposure time by roughly an order of magnitude longer with each increase in the correlation order.

An even more substantial caveat lies in the term $M^{1-\frac{R}{2}}$ since it shows that increasing the density of markers reduces the SNR of the $R$-order Q-ISM image for $R > 2$. The corruption of the image with a higher marker density is in stark contrast to our intuition regarding most microscopy methods such as confocal microscopy ($R = 1$) where a higher staining density is crucial for a high SNR. Moreover, fluorescent imaging of stained samples requires that the distance between markers be less than the resolution limit in order to reach the full resolution potential. Using an approximate $p = 0.01$ and a minimal value of $M = 10$ markers per diffraction limited spot, we arrive at the conclusion that reaching an SNR of 10 requires approximately a 50 second pixel dwell time, three orders of magnitude more than used in the imaging of fixed cells in this work. We conclude that, even under optimal conditions of current technology, imaging a biological sample with a higher correlation order seems impractical.

### Noise estimation in measured Q-ISM images

To reduce the impact of temporal fluctuations of the quantum dots, the scan that corresponds to Figure **2** in the main text was implemented by consecutively repeating a fast scan (50 $ms$ step dwell time) four times, to a total of 200 $ms$ acquisition time per scan step. Those repetitions were used for noise evaluation in the Q-ISM image. Figure *S6* compares the per-pixel standard deviation of those repeated scans with the predicted shot noise. Notably, the two noise evaluations are very similar, indicating shot noise as the dominant source of noise. As detailed above, the dominant noise source is that of shot noise in the number measured simultaneous photon pairs. The estimated error shown in (a) is therefore the square root of the sum over all detector pairs of a product of detection probabilities. The detection probabilities for each channel is estimated by dividing the number of detection per channel per scan step by the number of laser pulses.

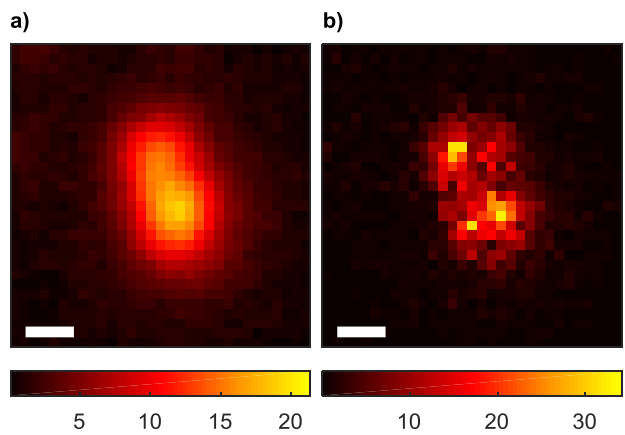

**Fig. S6| Q-ISM Noise assessment**. Two estimations for the noise in Fig.**2c** in the main text. **(a)** Poissonian noise, calculated as the square root of the an estimate for the number of simultaneous photon pairs at each pixel. **(b)** Standard deviation of the four scan repetitions that comprise the overall image in Fig.**2c**. Scale bar: $0.25\ \mu m$.

Supplementary section 7 – Theoretical description of Z-sectioning in Q-ISM

In this note we numerically calculate the Z-sectioning capabilities of Q-ISM and ISM, i.e. the dependence of the image total signal on the Z coordinate.

To express the Q-ISM signal dependency in the Z coordinate we write Eq.S9 with the full expression of the PSF in all three dimensions

$$\Delta G^{(2)}_{\alpha\beta}(x_s, Z_s) = \sum_{i=1}^{M} h_{exc}(x_i - x_s, Z_s) \cdot h_{img}(\bar{x}_a - (x_i - x_s), Z_s) \quad \text{(S40)}$$
$$\cdot h_{exc}(x_i - x_s, Z_s) \cdot h_{img}(\bar{x}_\beta - (x_i - x_s), Z_s)$$

where $x_s, x_i$ are two dimensional vectors representing the in-plane position of the scan coordinate and the markers. $\Delta G^{(2)}_{\alpha\beta}$ is the anti-buntching contrast as measured by two detectors positioned at $\bar{x}_a, \bar{x}_\beta$. The Z coordinate of the sample is denoted by $Z_s$. $h_{exc}$ and $h_{img}$ are the PSFs of excitation and detection respectively approximated here as a Gaussian beam

$$h_{exc}(x,z) = h_{img}(x,z) = \frac{2}{\pi \cdot w(z)^2} \cdot e^{-\frac{2|x|^2}{w(z)^2}}, \quad \text{(S41)}$$

where $w(z) = w_0 \sqrt{1 + \left(\frac{z}{z_R}\right)^2}$ is the beam's width as a function of z, $z_R = \frac{\pi \cdot w_0^2}{\lambda}$ is the Rayleigh range and $w_0$ is the beam's waist. Here we assume that the Stokes shift is negligible and therefore the parameters are equal for both the excitation and emission PSFs. To consider the contribution of different planes within a thick sample we calculate the integrated Q-ISM signal for a uniform sample. In this case, the image is independent of the scan's transverse position $x_s$ and therefore can be estimated in a single position. In an experimental realization of Q-ISM the integrated signal, measured by a finite detector array whose radius is denoted by $R$ (in the object plane), is given by

$$G^{(2)}_{ISM}(Z_s) = \int d^2 x_i \int_{r_\alpha < R} d^2 \bar{x}_a \int_{r_\beta < R} d^2 \bar{x}_\beta \cdot h_{exc}(x_i, Z_s) \cdot h_{img}(\bar{x}_a - x_i, Z_s) \quad \text{(S42)}$$
$$\cdot h_{exc}(x_i, Z_s) \cdot h_{img}(\bar{x}_\beta - x_i, Z_s)$$

Where the integration region for $x_i$ is over the entire plane $Z_s$ and the integration region for the detection coordinates, $\bar{x}_a$ and $\bar{x}_\beta$ is the detector's area. Using the expression written in Eq.S42 and integrating over the $x_i$ coordinate we obtain

$$G^{(2)}_{Q-ISM}(Z_s) \propto \frac{1}{w(z)^6} \cdot \int_{r_\alpha < R} d^2 \bar{x}_a \int_{r_\beta < R} d^2 \bar{x}_\beta \cdot e^{-\frac{1}{2w(z)^2}\left(3x_\alpha^2 + 3x_\beta^2 - 2x_\alpha \cdot x_\beta\right)} \quad \text{(S43)}$$

For a point detector $R \ll w(z)$ throughout the thickness of the sample the intergrand is approximately equal to one and the expression simplifies to

$$G^{(2)}_{Q-ISM}(Z_s) \propto \frac{R^4}{w(z)^6} \quad \text{(S44)}$$

The dependence of this expression of Z coordinate is presented in figure **S7** (black line). The contribution of out-of-focus planes decreases rapidly compared to a standard confocal microscope where the signal drops with the second power of $w(z)$. In order to evaluate the Q-

ISM signal dependence in the Z coordinate for our experimental setup, we calculate the integral given in Eq.S43 numerically for a detection array of radius $R \sim 250\ nm = 0.8 \cdot w_0$ (Fig **S7**. red line). We use here the estimated imaging PSF width (see Supplementary section 5) $w_0 = 309 \pm 5\ nm$ and an effective detector radius according to the total area of the array.

A similar analysis preformed for the ISM signal yield the expression

$$G_{ISM}^{(1)}(Z_s) \propto \frac{1}{w(z)^2} \cdot \int_{r_\alpha < R} d^2 \bar{x}_a \cdot e^{-\frac{x_\alpha^2}{w(z)^2}}. \tag{S45}$$

We numerically calculate the integral shown in Eq.S45 for the same experimental parameters as in the case of Q-ISM (Fig **S7**, blue line). The FWHM of the calculated Q-ISM and ISM signal are $720\ nm$ and $1470\ nm$ respectively.

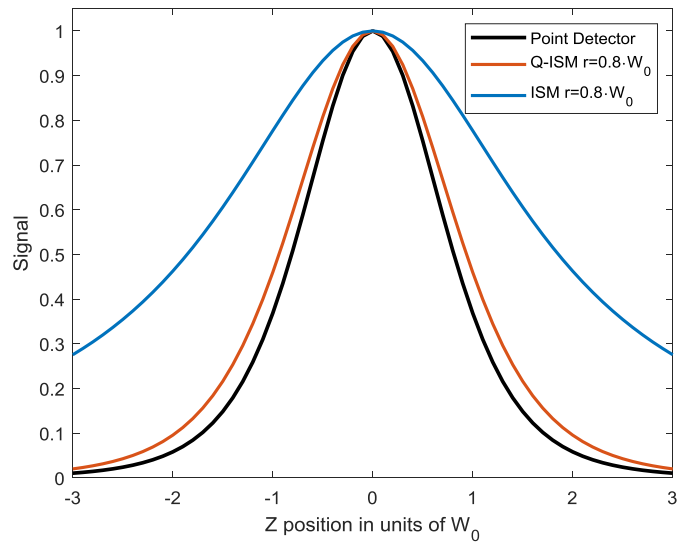

**Fig. S7| Calculated Z-sectioning of the ISM and Q-ISM methods.** The integrated signal for the Q-ISM method calculated for a point detector ($R \ll w(z)$) and the experimental setup in this work ($R \sim 250\ nm$) *versus* the axial coordinate Z are shown in black and red lines respectively. For the same experimental setup, we calculate the ISM signal *versus* axial coordinate z; shown in the blue line.

To examine the z-sectioning effect experimentally we measure the ISM and Q-ISM signal of a thin sample prepared by spin coating a dilute solution a solution of PMMA (3%wt) in Toluene with a dilute content of QDs. In a similar experiment to that presented in Figure 4 of the main text, we perform a confocal scan of the sample at different objective positions. The integrated values of ISM and Q-ISM signals obtained for each axial position are shown in figure **S8**. The FWHM of ISM and Q-ISM, calculated by linear interpolation, are $1.9\ \mu m$ and $0.8\ \mu m$ respectively. These values demonstrate that for our experimental conditions, measuring photon correlations enhances the z-sectioning resolution by $\sim 2.4$ with respect to standard ISM. The experimental values for ISM and Q-ISM Z-sectioning FWHMs are higher than the values calculated above by $\sim 20\%$ and $\sim 30\%$ respectively. This discrepancy is probably due to the over-simplified nature of the calculation, which approximates the high NA beam as a Gaussian beam and uses an effective detection area shape ignoring its pixelization, non-circular shape and gaps between fibers.

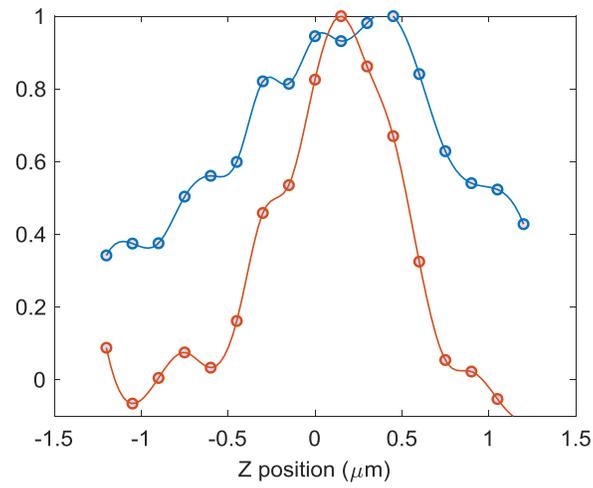

**Fig. S8| Estimation of Z-sectioning.** Blue and red lines present the integrated signal *versus* the sample height for the ISM and Q-ISM method respectively. Scans at different samples height were performed sequentially on a 2 $\mu m$ by 2 $\mu m$ area with 0.1 $\mu m$ scan steps and a 50 $ms$ pixel dwell time.

# References

Supplementary information references are included in the main text references above.